\documentclass[10pt,conference,english,twocolumn]{IEEEtran}
\IEEEoverridecommandlockouts

\pdfoutput=1

\usepackage{babel}
\usepackage[babel]{microtype}
\usepackage[babel]{csquotes}

\usepackage[svgnames]{xcolor}
\usepackage{graphicx}

\usepackage{amsmath}
\usepackage{amsfonts}
\usepackage{amssymb}
\usepackage{amsthm}
\usepackage{algorithm,algpseudocode}
\usepackage{bm}
\usepackage{siunitx}
\usepackage{booktabs}
\usepackage{subfigure}
\usepackage{tabularx}
\usepackage{titlesec} 
\usepackage{geometry}
 \usepackage[bookmarks=false,draft]{hyperref}

\usepackage{tikz}
\usepackage{pgfplots}
\pgfplotsset{compat=newest}
\usetikzlibrary{positioning}
\usetikzlibrary{shapes.geometric, arrows, decorations.pathreplacing}
\usetikzlibrary{shapes.arrows,arrows.meta}
\usetikzlibrary{3d}
\usetikzlibrary{calc}
\usetikzlibrary{external}

\usepackage[defernumbers=false,
backend=biber,
isbn=false,
url=false,
maxbibnames=999,
maxcitenames=10,
doi=false,
giveninits=true,
sortcites=true,
eprint=true,
hyperref=true,
style=ieee,
]{biblatex}
\bibliography{refs.bib}

\usepackage{hyperref}
\hypersetup{%
  pdftitle = {RISnet: A Domain-Knowledge Driven Neural Network Architecture for RIS Optimization with Mutual Coupling and Partial CSI},
  pdfauthor = {Bile Peng, Karl-Ludwig Besser, Shanpu Shen, Finn Siegismund-Poschmann, Ramprasad Raghunath, Daniel Mittleman, Vahid Jamali, Eduard A. Jorswieck},
  pdflang = {en-US},
  colorlinks,
  allcolors=.,
  urlcolor=MidnightBlue,
  bookmarksnumbered,
  bookmarksopen,
}

\usepackage[acronyms,nomain,xindy,nowarn]{glossaries}
\makeglossaries
\newacronym{isac}{ISAC}{integrated sensing and communication}
\newacronym{ae}{AE}{auto-encoder}
\newacronym{vae}{VAE}{variational auto-encoder}
\newacronym{ap}{AP}{access point}
\newacronym{aoa}{AoA}{angle-of-arrival}
\newacronym{ao}{AO}{alternating optimization}
\newacronym{bs}{BS}{base station}
\newacronym{dnn}{DNN}{deep neural network}
\newacronym{cnn}{CNN}{convolutional neural network}
\newacronym{rnn}{RNN}{recurrent neural network}
\newacronym{ma}{MA}{multiple access}
\newacronym{nn}{NN}{neural network}
\newacronym{ue}{UE}{user equipment}
\newacronym{noma}{NOMA}{non-orthogonal multiple access}
\newacronym{mimo}{MIMO}{multiple-input-multiple-output}
\newacronym{miso}{MISO}{multiple-input-single-output}
\newacronym{cf}{CF mMIMO}{cell-free massive MIMO}
\newacronym{cdf}{CDF}{cumulative distribution function}
\newacronym{gnn}{GNN}{graph neural network}
\newacronym{cc}{CC}{channel charting}
\newacronym{los}{LoS}{line-of-sight}
\newacronym{csi}{CSI}{channel state information}
\newacronym{star}{STAR}{simultaneously transmitting and reflecting}
\newacronym{aod}{AoD}{angle-of-departure}
\newacronym{lwf}{LwF}{learing without forgetting}
\newacronym{sinr}{SINR}{signal-to-interference-plus-noise ratio}
\newacronym{snr}{SNR}{signal-to-noise ratio}
\newacronym{sgd}{SGD}{stochastic gradient descent}
\newacronym{agi}{AGI}{artificial general intelligence}
\newacronym{llm}{LLM}{large language model}
\newacronym{dl}{DL}{deep learning}
\newacronym{rl}{RL}{reinforcement learning}
\newacronym{crb}{CRB}{Cram\'er-Rao bound}
\newacronym{tpu}{TPU}{tensor processing unit}
\newacronym{evt}{EVT}{extreme value theory}
\newacronym{urllc}{URLLC}{ultra-reliable low latency communications}
\newacronym{mrt}{MRT}{maximum ratio transmission}
\newacronym{embb}{eMBB}{enhanced mobile broadband}
\newacronym{mmse}{MMSE}{minimum mean square error}
\newacronym{qd}{QD}{quasi degradation}
\newacronym{rss}{RSS}{received signal strength}
\newacronym{siso}{SISO}{single-input single-output}
\newacronym{rsma}{RSMA}{rate-splitting multiple access}
\newacronym{lstm}{LSTM}{long short-term memory}
\newacronym{drl}{DRL}{deep reinforcement learning}
\newacronym{dqn}{DQN}{deep Q-network}
\newacronym{ml}{ML}{machine learning}
\newacronym{sca}{SCA}{successive convex approximation}
\newacronym{admm}{ADMM}{alternating direction method of multipliers}
\newacronym{mm}{MM}{majorization-minimization}
\newacronym{rmcg}{RMCG}{Riemannian manifold conjugate gradient}
\newacronym{sdma}{SDMA}{space-division multiple access}
\newacronym{sdr}{SDR}{semidefinite relaxation}
\newacronym{mse}{MSE}{mean squared error}
\newacronym{ris}{RIS}{reconfigurable intelligent surface}
\newacronym{tof}{ToF}{time of flight}
\newacronym{em}{EM}{expectation-maximization}
\newacronym{zf}{ZF}{zero-forcing}
\newacronym{ekf}{EKF}{extended Kalman filter}
\newacronym{bcd}{BCD}{block coordinate descent}
\newacronym{concrete}{CONCRETE}{CONtinuous relaxation for disCRETE}
\newacronym{mop}{MOP}{multi-objective programming}
\newacronym{uwb}{UWB}{ultra-wide band}
\newacronym{sic}{SIC}{successive interference cancellation}
\newacronym{otfs}{OTFS}{orthogonal time frequency space}
\newacronym{ofdm}{OFDM}{orthogonal frequency-division multiplexing}
\newacronym{mpc}{MPC}{multi-path component}
\newacronym{wmmse}{WMMSE}{weighted minimum mean square error}
\newacronym{wsr}{WSR}{weighted sum-rate}
\newacronym{mab}{MAB}{multi-armed bandit}
\newacronym{ppo}{PPO}{proximal policy optimization}
\newacronym{sac}{SAC}{soft actor-critic}
\newacronym{iid}{i.i.d.}{independent and identically distributed}
\newacronym{wrt}{w.r.t.}{with respect to}
\newacronym{fdma}{FDMA}{frequency domain multiple access}
\newacronym{tdma}{TDMA}{time domain multiple access}
\newacronym{cdma}{CDMA}{code domain multiple access}
\setacronymstyle{long-short}
\glsdisablehyper

\addto\extrasenglish{%

}
\addto\extrasenglish{%

}

\IfPackageLoadedTF{titlesec}{%
	\ifCLASSOPTIONconference%
	\titlespacing{\section}{0pt}{1.5ex plus 1.5ex minus 0.5ex}{0.7ex plus 1ex minus 0ex} 
	\titlespacing{\subsection}{0pt}{1.5ex plus 1.5ex minus 0.5ex}{0.7ex plus .5ex minus 0ex} 
	\else
	\titlespacing{\section}{0pt}{3.0ex plus 1.5ex minus 1.5ex}{0.7ex plus 1ex minus 0ex} 
	\titlespacing{\subsection}{0pt}{3.5ex plus 1.5ex minus 1.5ex}{0.7ex plus .5ex minus 0ex} 
	\fi
	
	\def\thesubsubsectiondis{\arabic{subsubsection})}
	\def\theparagraphdis{\alph{paragraph})}
	\titleformat{\subsubsection}[runin]{\normalfont\normalsize\itshape}{\thesubsubsectiondis}{.5em}{}[:]
	\titlespacing*{\subsubsection}{\parindent}{0ex plus 0.1ex minus 0.1ex}{1ex}
	\titleformat{\paragraph}[runin]{\normalfont\normalsize\itshape}{\theparagraphdis}{.5em}{}[:]
	\titlespacing*{\paragraph}{2\parindent}{0ex plus 0.1ex minus 0.1ex}{1ex}
}{}

\title{
A Scalable Machine Learning Approach Enabled RIS Optimization with Implicit Channel Estimation
}
\author{
\IEEEauthorblockN{
Bile Peng\IEEEauthorrefmark{1}, 
Vahid Jamali\IEEEauthorrefmark{2}, 
and Eduard Jorswieck\IEEEauthorrefmark{1}
}
\IEEEauthorblockA{\IEEEauthorrefmark{1}Institute for Communications Technology, Technische Universit\"at Braunschweig, Germany}
\IEEEauthorblockA{\IEEEauthorrefmark{2} Resilient Communication Systems Lab, Technische Universit\"at Darmstadt, Germany}
Email: \{b.peng, e.jorswieck\}@tu-braunschweig.de, vahid.jamali@tu-darmstadt.de
\thanks{The work of B.~Peng and E.~Jorswieck is supported by the Federal Minist of Research, Technology and Space (BMFTR), Germany, as part of the 6G Research and Innovation Cluster (6G-RIC) under Grant 16KISK031.
The work of B.~Peng is supported by the German research foundation (DFG) as part of the ML4RIS project (566937681).
The work of Vahid Jamali was supported in part by the LOEWE Initiative, Hesse, Germany, within
the emergenCITY Center [LOEWE/1/12/519/03/05.001(0016)/72].
The work of E. Jorswieck received partly support from the Smart Networks and Services Joint Undertaking (SNS JU) under the
European Union’s Horizon Europe research and innovation programme within 6G-SENSES project (Grant Agreement No 101139282).
}
}

\begin{document}

\maketitle

\begin{abstract}
The \gls{ris} is considered as a key enabler of the next-generation mobile radio systems.
While attracting extensive interest from academia and industry due to its
passive nature and low cost,
scalability of \gls{ris} elements and requirement for \gls{csi} 
are two major difficulties for the \gls{ris} to become a reality.
In this work,
we introduce an unsupervised \gls{ml} enabled optimization approach to configure the \gls{ris}.
The dedicated \gls{nn} architecture RISnet is combined with an implicit channel estimation method.
The RISnet learns to map from received pilot signals to \gls{ris} configuration directly without explicit channel estimation.
Simulation results show that the proposed algorithm outperforms baselines significantly.
\end{abstract}

\begin{IEEEkeywords}
Reconfigurable intelligent surface,
machine learning,
implicit channel estimation.
\end{IEEEkeywords}

\glsresetall
\section{Introduction}
\label{sec:intro}

The \gls{ris} is widely investigated as a key technology of 6G
because of its ability to dynamically optimize the radio propagation channel. 
By manipulating the electromagnetic waves in a passive manner, 
\gls{ris} can shape the environment to enhance the communication system performance. 
Its ability to tailor the wireless environment to the needs of different devices and applications is
fundamental to realizing the high-performance demands of future 6G networks~\cite{liu2021reconfigurable}.

In order to realize \gls{ris} in reality,
two main challenges must be addressed: 
\emph{scalability} to a large number of \gls{ris} elements 
and \emph{channel estimation}. 
A large number of \gls{ris} elements is necessary to achieve a sufficient received signal strength
due to the passive nature of the \gls{ris}.
However,
the large number of elements poses a difficult optimization problem.
This scalability issue also impacts the second challenge, 
namely, channel estimation, 
because \gls{csi} is essential for configuring the \gls{ris} effectively. 
With more \gls{ris} elements, 
the complexity of estimating the channels for each element grows significantly, 
leading to computational and operational challenges. 

In the literature,
various algorithms have been proposed to optimize \gls{bs} precoding and \gls{ris} configuration jointly.
For \gls{sdma},
precoding schemes include
\gls{mrt}, \gls{zf}, \gls{mmse} precoding~\cite{joham2005linear},
and \gls{wmmse} precoding with proved equivalence to \gls{wsr} maximization~\cite{shi2011iteratively}.
The optimality of \gls{noma} in degraded channels was demonstrated in~\cite{jorswieck2021optimality}.
An optimal \gls{noma} precoding scheme for a multi-antenna \gls{bs} in a quasi-degraded channel was derived in~\cite{chen2016optimal,chen2016application}.
The joint optimization of precoding and \gls{ris} configuration was performed with
 \gls{bcd}~\cite{guo2020weighted},
\gls{mm}~\cite{huang2018achievable,zhou2020intelligent} and
\gls{admm}~\cite{liu2021two} algorithms to maximize the \gls{wsr} in \gls{sdma}.
The \gls{ris} was also applied to make a channel quasi-degraded and minimize the transmit power subject to the required rate using \gls{sca}~\cite{zhu2020power} and
\gls{sdr}~\cite{fu2019intelligent,yang2021reconfigurable} algorithms.
In addition,
\gls{rmcg} and Lagrangian method were applied to optimize multiple \glspl{ris} and \glspl{bs} to serve cell-edge users~\cite{li2020weighted}.
The impact of \glspl{ris} on the outage probability in \gls{noma} was studied in~\cite{hou2020reconfigurable}.
Successive refinement algorithm and exhaustive search were applied for passive beamforming improvement~\cite{wu2019beamforming}.
The active \glspl{ris} was optimized with the \gls{sca} algorithm to maximize the \gls{snr}~\cite{long2021active}.
The gradient-based optimization was applied to optimize the effective rank and the minimum singular value~\cite{Elmossallamy2021spatial}.

In general,
the above analytical iterative methods do not scale well with the number of \gls{ris} elements.
Most works assume no more than 100 elements, 
which is far from the vision of more than 1000~\gls{ris} elements~\cite{di2020smart}
and the requirement in many scenarios to realize a necessary link budget~\cite{najafi2020physics}.
Moreover, the required numbers of iterations make the proposed iterative algorithms difficult to be implemented in real time
since the computation time is longer than the channel coherence time.
Furthermore,
even if the algorithm has a high scalability to  optimize more than 1000 elements,
such a large number makes the \gls{csi} extremely difficult to obtain.
Therefore, the assumption of known \gls{csi} is very unrealistic.

A noticeable effort is to apply \gls{ml} to optimize the \gls{ris},
which bypasses the difficulty of analytical solution via the universal approximation property of the \gls{nn}~\cite{hornik1989multilayer}.
Recently, \gls{dl} and \gls{rl} were applied and compared for \gls{ris} optimization~\cite{zhong2021ai}.
\Gls{lstm} and \gls{dqn} were applied to optimize \gls{ris} for \gls{noma}~\cite{gao2021machine}.
The \gls{star}-\gls{ris} was combined with \gls{noma}~\cite{yue2022simultaneously}.
\Gls{rl} was applied to maximize the sum-rate in
\gls{sdma}~\cite{huang2020reconfigurable}. 
The achievable rate was predicted and the \gls{ris} was configured with \gls{dl}~\cite{sheen2021deep}.
The \gls{ris} was configured directly with received pilot signals~\cite{jiang2021learning}.
The mapping from received pilot signal to the phase shifts was optimized~\cite{ozdougan2020deep}.
Due to the separation of training and testing phases,
the trained \gls{ml} model was able to be applied in real time.
However, the scalability with the number of \gls{ris} elements was still limited
and the assumption of known \gls{csi} is unrealistic given many \gls{ris} elements.

In this work,
we propose an unsupervised \gls{ml} method to autonomously optimize the \gls{ris} configuration.
It combines the dedicated \gls{nn} architecture RISnet~\cite{peng2025risnet} and
an implicit channel estimation.
In this way, we realize scalability and does not require explicit channel estimation.
Simulation results show that the proposed RISnet can configure an \gls{ris} with 1296 elements
without explicit channel estimation
to outperform baselines in both performance and computation time.
\section{Problem Statement}
\label{sec:problem}

\subsection{System Model}

\begin{figure}[htbp]
    \centering
    \resizebox{.8\linewidth}{!}{
    \begin{tikzpicture}
    \tikzstyle{base}=[isosceles triangle, draw, rotate=90, fill=gray!60, minimum size =.5cm]
	\tikzstyle{user}=[rectangle, draw, fill=gray!60, minimum size =.5cm, rounded corners=0.1cm]
	\tikzstyle{element}=[rectangle, fill=gray!30]
	
	\node[base,label={left:BS}] (BS) at (-3,0){};
	\draw[decoration={expanding waves,segment length=6},decorate] (BS) -- (-3,1.5);
	\node[user,label={below:User 1}] (UE1) at (4,2){};
	\node[user,label={below:User $U$}] (UE2) at (4,0){};
	\draw[step=0.33cm,thick] (-1,1.98) grid (0, 3);
	\node[label={[label distance=10]above:RIS}] (RIS) at (-0.5, 2.5) {};
	
	\node at ($(UE1)!.5!(UE2)$) {\vdots};
	\node[above right=.15 and .3 of RIS]{$\boldsymbol{\Phi}$ (\gls{ris} signal processing)};
	
	\draw[-to,shorten >=3pt] (BS) to node[above left, pos=.1, yshift=-1.6cm] {$\mathbf{V}$ (precoding)} (UE1);
	\draw[-to,shorten >=3pt] (BS) to node[above,pos=.3, ] {$\mathbf{D}$} (UE2);
	
	\draw[-to] (BS) to node[above left, pos=.6] {$\mathbf{H}$} (RIS);
	
	\draw[-to,shorten >=3pt] (RIS) to node[left=-1mm, below=0mm, pos=.3] {$\mathbf{G}$} (UE1);
	\draw[-to,shorten >=3pt] (RIS) to node[below=-3mm, left=2mm] {} (UE2);
\end{tikzpicture}}
    \caption{The system model. The RIS manipulates the wireless channel property to assist the BS,
    which serves multiple users with SDMA.}
    \label{fig:system_model}
\end{figure}
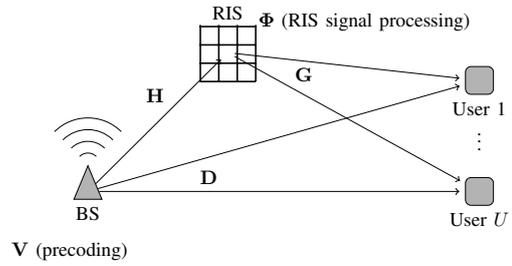

We consider a downlink multi-user scenario,
as shown in \autoref{fig:system_model}.
The \gls{bs} has multiple antennas whereas
each user is equipped with a single antenna.
We assume that the \gls{bs} serves the users with the same radio resource block using \gls{sdma}.
We denote the channel from \gls{bs} to \gls{ris} as $\mathbf{H} \in \mathbb{C}^{N\times M}$,
where $N$ is the number of \gls{ris} elements
and $M$ is the number of \gls{bs} antennas,
the channel from \gls{ris} to users as $\mathbf{G} \in \mathbb{C}^{U\times N}$,
where $U$ is the number of users.
The signal processing by the \gls{ris} is described by the diagonal matrix
$\boldsymbol{\Phi} \in \mathbb{C}^{N \times N}$
with element
in row~$n$ and column~$n$ is defined as
$\phi_{nn}=e^{j\pi \varphi_n}$ 
and
$\varphi_n$ is the phase shift of \gls{ris} element~$n$.
Note that $|\phi_{nn}|=1$,
indicating the passive nature of the \gls{ris}.
The direct channel from \gls{bs} to users directly without \gls{ris} is denoted as $\mathbf{D} \in \mathbb{C}^{U\times M}$.

In the \gls{ris}-assisted downlink transmission,
the end-to-end channel $\mathbf{C} \in \mathbb{C}^{U \times M}$ between \gls{bs} and users is described by
\begin{equation}
    \mathbf{C} = \mathbf{D} + \mathbf{G} \boldsymbol{\Phi}\mathbf{H}.
    \label{eq:reduced_channel}
\end{equation}
The received signal by the users is
\begin{equation}
\mathbf{y} =  \mathbf{C}
\mathbf{V} \mathbf{x} + \mathbf{n},
\label{eq:transmission_los}
\end{equation}
where $\mathbf{x} \in \mathbb{C}^{U \times 1}$ is the transmitted symbols,
$\mathbf{V} \in \mathbb{C}^{M \times U}$ is the precoding matrix,
$\mathbf{y} \in \mathbb{C}^{U \times 1}$ is the received symbols, 
and $\mathbf{n} \in \mathbb{C}^{U \times 1}$ is the noise. 
Our objective is to maximize the \gls{wsr}
of all users 
subject to the maximum available power constraint,
i.e.,

\begin{subequations}
\begin{align}
    \max_{\mathbf{V}, \boldsymbol{\Phi}}\quad & 
  f=\sum_{u=1}^U w_u\log_2\left(1+\frac{|c_{uu}|^2}{\sum_{v\neq u}|c_{uv}|^2+\sigma^2}\right) \label{eq:sdma_objective}
\\
    \text{s.t.}\quad & \text{Tr}\left(\mathbf{V}\mathbf{V}^H\right) \leq E_{Tr}, \label{eq:sdma_transmit_power}\\
    & |\phi_{nn}|=1 \text{ for } n=1, \dots, N, \label{eq:sdma_passive_ris}\\
    & |\phi_{nn'}|=0 \text{ for } n, n' = 1, \dots, N \text{ and }  n \neq n', \label{eq:sdma_offdiagonal}
\end{align}
\label{eq:problem_sdma}%
\end{subequations}
where $c_{uv}$ is the element in row~$u$ and column~$v$ of $\mathbf{C}$,
and $w_u$ is the weight of user~$u$.
Constraint~\eqref{eq:sdma_transmit_power} is the maximum available power constraint,
Constraint~\eqref{eq:sdma_passive_ris} specifies that
the \gls{ris} is a passive device and cannot amplify the received signal,
Constraint~\eqref{eq:sdma_offdiagonal}
indicates that there is no energy transfer between the \gls{ris} elements.

In order to realize a good performance with
realistic assumptions,
it is desirable to have a large \gls{ris} with many elements,
such that a high channel gain can be realized
despite \gls{ris}' passive nature.
However,
the scalability with \gls{ris} elements is a major challenge of algorithm design.
Furthermore,
the large number of \gls{ris} elements also poses a severe challenge of channel estimation.
The assumption of known full \gls{csi} is unrealistic in reality.
In the following sections,
we propose an \gls{ml}-enabled solution to Problem \eqref{eq:problem_sdma}
with high scalability to more than one thousand \gls{ris} elements
and without assumption of known \gls{csi}.
\section{Proposed Machine Learning Enabled Solution}
\label{sec:ml}

\subsection{Framework of Unsupervised Machine Learning for Optimization}
\label{sec:ml_framework}

While the mainstream approach in \gls{ml} is supervised learning~\cite{yu2022role}, 
we propose an unsupervised learning scheme.
This approach is motivated by the lack of known optimal \gls{ris} configurations
for complex problem instants, 
allowing the algorithm to autonomously discover the best solution without labels provided by human.
Its framework is presented as follows:
given a problem representation $\boldsymbol{\Gamma}$
(in Problem \eqref{eq:problem_sdma}, the information about channels
and the user weights),
we look for a solution $\boldsymbol{\Psi}$
(\gls{bs} precoding $\mathbf{V}$ and \gls{ris} configuration $\boldsymbol{\Phi}$)
that maximizes objective $f$  \eqref{eq:sdma_objective},
which is fully determined by $\boldsymbol{\Gamma}$ and $\boldsymbol{\Psi}$.
Therefore, it can be written as
$f(\boldsymbol{\Gamma}, \boldsymbol{\Psi})$.
We define an \gls{nn} $N_\theta$,
where $\theta$ is the trainable parameters,
i.e., weights and biases in all layers of $N_\theta$.
$N_\theta$
maps from $\boldsymbol{\Gamma}$
to $\boldsymbol{\Psi}$,
i.e., $\boldsymbol{\Psi} = N_\theta(\boldsymbol{\Gamma})$.
We write the objective as $ f(\boldsymbol{\Gamma}, \boldsymbol{\Psi})=f(\boldsymbol{\Gamma}, N_\theta(\boldsymbol{\Gamma}); \theta).$
Note that it is emphasized that $f$ depends on $\theta$.
We then collect massive data of problem representations in a training set $\mathcal{D}$
and formulate the problem as
\begin{equation}
    \min_\theta K(\theta)=\sum_{\boldsymbol{\Gamma} \in \mathcal{D}} f(\boldsymbol{\Gamma}, N_\theta(\boldsymbol{\Gamma}); \theta),
    \label{eq:ml}
\end{equation}
where $K(\theta)$ is the data-driven objective function.
In this way, $N_\theta$ is optimized for the emsemble of $\boldsymbol{\Gamma} \in \mathcal{D}$
(\emph{training}) using gradient decent:
\begin{equation}
    \theta \leftarrow \theta - \eta \nabla_\theta K(\theta),
    \label{eq:gradient-ascend}
\end{equation}
where $\eta$ is the learning rate.
If $N_\theta$ is well trained,
$\boldsymbol{\Psi}' = N_\theta (\mathbf{\Gamma'})$ is also a good solution for $\boldsymbol{\Gamma}' \notin \mathcal{D}$ (\emph{testing}).
This is similar to that a human uses experience to solve new problems of the same type.

In particular,
$\boldsymbol{\Gamma}$ is difficult to obtain in some problems,
e.g., due to high radio resource requirement for the \gls{csi} acquisition,
we use $\mathbf{O}$ as an \emph{observation} of $\boldsymbol{\Gamma}$ to replace $\boldsymbol{\Gamma}$ as the input of $N_\theta$.
If $\mathbf{O}$ contains sufficient information about $\boldsymbol{\Gamma}$ and it is much easier to obtain than $\boldsymbol{\Gamma}$,
using $\mathbf{O}$ as the input of $N_\theta$ can significantly improve the feasibility of \gls{ml} enabled optimization.

\subsection{Implicit Channel Estimation}
\label{sec:implicit}

Conventionally,
the problem representation $\boldsymbol{\Gamma}$ is defined based on the \gls{csi}~\cite{zhong2021ai,peng2025risnet}
and consequently,
the \gls{ris} configuration is computed according to the \gls{csi}.
However,
this approach assumes known \gls{csi},
which is often unrealistic 
due to the large number of \gls{ris} elements.
Therefore,
we propose an \gls{ris} optimization scheme,
which maps from received pilot signals from users $\mathbf{O}$ directly to \gls{ris} configuration,
such that no assumption of known \gls{csi} is required.

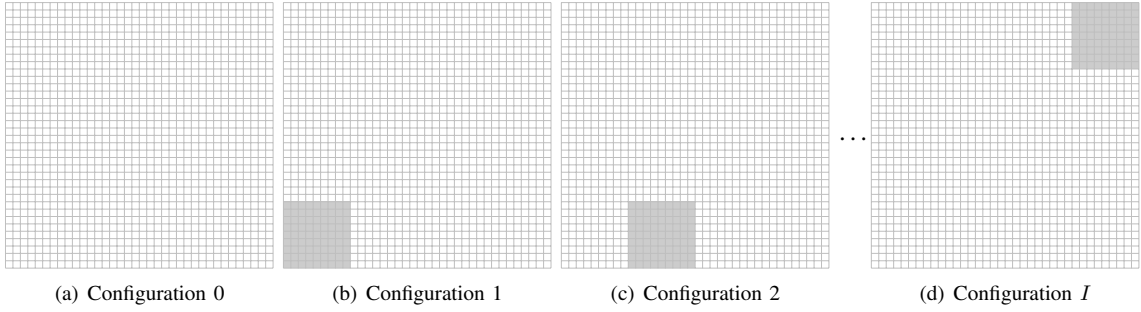
\begin{figure*}[htbp]
    \centering
    \subfigure[Configuration 0\label{fig:config0}]{\resizebox{.2\linewidth}{!}{\begin{tikzpicture}

\draw[step=1cm, gray!50!white] (0,0) grid (36, 36);
\end{tikzpicture}}}
    \subfigure[Configuration 1\label{fig:config1}]{\resizebox{.2\linewidth}{!}{\begin{tikzpicture}
\fill[black!20!white] (0, 0) rectangle (9, 9);

\draw[step=1cm, gray!50!white] (0,0) grid (36, 36);
\end{tikzpicture}}}
    \subfigure[Configuration 2\label{fig:config2}]{\resizebox{.2\linewidth}{!}{\begin{tikzpicture}
\fill[black!20!white] (9, 0) rectangle (18, 9);

\draw[step=1cm, gray!50!white] (0,0) grid (36, 36);
\end{tikzpicture}}}
    \raisebox{17mm}\dots
    \subfigure[Configuration $I$\label{fig:configI}]
    {\resizebox{.2\linewidth}{!}{\begin{tikzpicture}
\fill[black!20!white] (27, 27) rectangle (36, 36);

\draw[step=1cm, gray!50!white] (0,0) grid (36, 36);
\end{tikzpicture}}}
    \caption{RIS configurations for implicit channel estimation.
    White elements have a phase shift of 0
    where gray elements have a phase shift of $\pi$.}
    \label{fig:configuration}
\end{figure*}

In our proposed approach,
users send orthogonal pilot signals via both \gls{ris}
and direct channel to \gls{bs} repeatedly multiple time steps.
For each time step,
the \gls{ris} has a different configuration,
such that the received pilot signals are different
although the pilot signals are identical
and the channels are assumed static.
As shown in \autoref{fig:configuration},
in time step~0,
all \gls{ris} elements have a phase shift of 0,
i.e.,
\gls{ris} configuration $\boldsymbol{\Phi}(0)=\mathbf{I}$,
as depicted in \autoref{fig:config0}.
In time step~$i=1, \dots, I$,
where $I + 1$ is the total number of \gls{ris} configurations for channel estimation,
a subset of \gls{ris} element $\Lambda_i$ has a phase shift of $\pi$,
all other elements have a phase shift of 0,
i.e., $\phi_{nn}=-1$ for $n \in \Lambda_i$ and 
$\phi_{nn}=1$ for $n \notin \Lambda_i$.
$\{\Lambda_i\}_{i=1, \dots, I}$ is defined in such a way
that $\Lambda_i (i=1, \dots, I)$ are disjoint subsets with union equal to set of all \gls{ris} elements, i.e.,
\begin{equation}
    \bigcup_{i=1, \dots, I} \Lambda_i = \{1, \dots, N\},
    \label{eq:union}
\end{equation}
and
\begin{equation}
    \Lambda_i \cap \Lambda_j = \emptyset \text{ for } i \neq j.
    \label{eq:intersection}
\end{equation}

As an example,
we depict an \gls{ris} array of shape $36\times 36$
in \autoref{fig:configuration}.
The \gls{ris} is divided into 16 blocks of shape $9\times 9$.
In configuration~$i > 0$,
\gls{ris} elements in the $i$th block have a phase shift of $\pi$,
as shown in \autoref{fig:config1} - \autoref{fig:configI}.
Denote the pilot signal of user~$u$ as $s_u$,
and the received pilot signal from user~$u$ at time step~$i$ as $\mathbf{y}_{u}(i)$,
we have
\begin{equation}
    \mathbf{y}_{u}(i) = (\mathbf{d}_u + \mathbf{H}^T \boldsymbol{\Phi}(i) \mathbf{g}_u) s_u + \mathbf{n}_{u}(i),
    \label{eq:received_pilot_signal}
\end{equation}
where $\mathbf{d}_u$ is the direct channel from user~$u$ to \gls{bs},
i.e.,
$\mathbf{d}_u \in \mathbb{C}^{M\times 1}$ is the transposed $u$th row of $\mathbf{D}$,
$\mathbf{g}_u \in \mathbb{C}^{N \times 1}$ is the channel from user~$u$ to \gls{ris},
i.e., $\mathbf{g}_k$ is the transposed $k$th row of $\mathbf{G}$,
$\mathbf{n}_{ki} \in \mathbb{C}^{M\times 1}$ is the thermal noise.
Note that \eqref{eq:received_pilot_signal} is the signal transmission in uplink whereas
\eqref{eq:reduced_channel} is in downlink.
Therefore, the order of the \gls{ris}-assisted cascaded channel is reversed
and the channel matrices are transposed.
Since the difference between $\boldsymbol{\Phi}(0)$ and $\boldsymbol{\Phi}(i)$ ($i > 0$) is only elements in $\Lambda_i$,
we have
\begin{equation}
    \mathbb{E}\left(\mathbf{y}_{u}(i) - \mathbf{y}_{k0}\right) = 
    \sum_{n \in \Lambda_i}2\mathbf{h}_n g_{un} s_u,
    \label{eq:received_pilot_diff}
\end{equation}
where
$\mathbb{E}$ denotes the expectation operator with respect to the random thermal noise,
$\mathbf{h}_n \in \mathbb{C}^{M \times 1}$ is the channel from \gls{ris} element~$n$ to \gls{bs},
i.e., the $n$th row of $\mathbf{H}$ transposed,
$g_{un}$ is the channel from user~$u$ to \gls{ris} element~$n$,
i.e., the element in row~$k$ and column~$n$ of $\mathbf{G}$.
From \eqref{eq:received_pilot_diff},
we note that the difference between received pilot signal $s_u$ from user~$u$
using \gls{ris} configurations 0 and $i$ ($i > 0$)
contains information about channel $g_{un}, n \in \Lambda_i$.
We use
$\mathbf{y}_{u}(i) - \mathbf{y}_{u}(0)$
as a noisy observation of $\mathbb{E}\left(\mathbf{y}_{u}(i) - \mathbf{y}_{u}(0)\right)$ in \eqref{eq:received_pilot_diff}.
The concatenation of $\{\mathbf{y}_{u}(i) - \mathbf{y}_{u}(0)\}$ for all $u=1, \dots, U$ and $i=1, \dots, I$ is used as the input of the \gls{nn} as channel features of the cascaded channel $\mathbf{H}\boldsymbol{\Phi}\mathbf{G}$.
For the direct channel $\mathbf{D}$,
note from \eqref{eq:union} and \eqref{eq:intersection} that
\begin{equation}
    (I-2)\boldsymbol{\Phi}(0) - \sum_{i=1}^I \boldsymbol{\Phi}(i) = \mathbf{0}.
\end{equation}
Therefore, we have
\begin{equation}
    \mathbb{E}\left((I-2)\mathbf{y}_{u}(0) - \sum_{i=1}^I \mathbf{y}_{u}(i) \right) = -2\mathbf{d}_u s_u,
    \label{eq:direct_channel_estimation}
\end{equation}
i.e., $(I-2)\mathbf{y}_{u}(0) - \sum_{i=1}^I \mathbf{y}_{u}(i)$ contains information
about the direct channel $\mathbf{d}_u$.
We use $(I-2)\mathbf{y}_{u}(0) - \sum_{i=1}^I \mathbf{y}_{u}(i)$ 
as a noisy observation of $\mathbb{E}\left((I-2)\mathbf{y}_{u}(0) - \sum_{i=1}^I \mathbf{y}_{u}(i) \right)$.
The concatenation of $\{ (I-2)\mathbf{y}_{u}(0) - \sum_{i=1}^I \mathbf{y}_{u}(i) \}$
for $u=1, \dots, U$ is used as the input of the \gls{nn}
as channel features of the direct channel $\mathbf{D}$.

\subsection{Proposed RISnet Architecture}
\label{sec:risnet}

We introduce a dedicated \gls{nn} architecture RISnet with implicit channel estimation.
We do not choose an existing \gls{nn} architecture but design a new \gls{nn} architecture due to the following reasons:
\begin{itemize}
\item Scalability: Conventionally, the \gls{nn} complexity grows with the dimensions of \gls{nn} input and output.
However, we aim to develop a new architecture that enables configuration of a large \gls{ris}
with a low-complexity \gls{nn} network.
\item Mapping from received pilot signals to \gls{ris} phase shifts:
The observation $\mathbf{O}$ is a partial observation of $\boldsymbol{\Gamma}$.
The observed $\{\mathbf{y}_{u}(i) - \mathbf{y}_{u}(0)\}$ contains the information about the sum $\sum_{n\in \Lambda_i} 2\mathbf{h}_ng_{un}s_u$.
Based on it,
we determine individual $\phi_n$ for $n \in \Lambda_i$.
This relationship between input and output must be reflected in the \gls{nn} architecture design.
\end{itemize}

Motivated by the facts that
the phase shift $\phi_n$ of \gls{ris} element~$n$
depends on the \gls{csi} of itself
as well as a common optimization objective shared by all \gls{ris} elements,
and the data rate is determined by signal strength and interference of other users,
we define four categories of features
for each \gls{ris} element:
\begin{itemize}
    \item Category \texttt{cc} for \texttt{c}urrent user and \texttt{c}urrent element
    \item Category \texttt{ca} for \texttt{c}urrent user and \texttt{a}ll elements
    \item Category \texttt{oc} for \texttt{o}ther users and \texttt{c}urrent element
    \item Category \texttt{oa} for \texttt{o}ther users and \texttt{a}ll elements.
\end{itemize}
The inference of an intermediate layer of RISnet is performed as
\begin{multline}
\mathbf{f}_{un, i + 1} =\\
\begin{pmatrix}
     \text{ReLU}(\mathbf{W}^{\texttt{cc}}_{i} \mathbf{f}_{un, i} + \mathbf{b}_i^{\texttt{cc}}) \\
     \left(\sum_{n'}\text{ReLU}(\mathbf{W}^{\texttt{ca}}_{i} \mathbf{f}_{un', i} + \mathbf{b}_i^{\texttt{ca}})\right) \big/ N\\
     \left(\sum_{u'\neq u}\text{ReLU}(\mathbf{W}^{\texttt{oc}}_{i} \mathbf{f}_{u'n, i} + \mathbf{b}_i^{\texttt{oc}})\right) \big/ (U-1)\\
     \left(\sum_{u'\neq u}\sum_{n'}\text{ReLU}(\mathbf{W}^{\texttt{oa}}_{i} \mathbf{f}_{u'n', i} + \mathbf{b}_i^{\texttt{oa}})\right) \big/ (N(U-1)) 
\end{pmatrix},
\label{eq:layer_processing}
\end{multline}
where the four rows on the right hand side correspond to the above-described four categories,
$\mathbf{W}_i^{\texttt{cc}}$ is the trainable weight matrix of category~\texttt{cc} on layer~$i$,
$\mathbf{f}_{un,i}$ is the input feature vector of user~$u$ and \gls{ris} element~$n$ on layer~$i$,
$\mathbf{b}_i^\texttt{cc}$ is the bias vector of category~\texttt{cc} on layer~$i$.
We observe from the first row of the right hand side of \eqref{eq:layer_processing} that
the information processing is like a conventional fully connected layer but
is applied to the feature vector
of one user and one \gls{ris} element,
rather than the feature vector of all users and \gls{ris} elements.
For the other categories,
the outputs per user and \gls{ris} element are averaged over other users and/or all \gls{ris} elements
as a common context of users/\gls{ris} elements.

The inference \eqref{eq:layer_processing} produces feature of one \gls{ris} element and one user.
Since the input is not observation of single channel gain of each \gls{ris} element,
it is required to \emph{unwrap} the observed sum of channel gains to features of individual \gls{ris} elements.
This process is done as
\begin{multline}
\mathbf{f}_{u\nu(n, j), i + 1} = \\
\begin{pmatrix}
     \text{ReLU}(\mathbf{W}^{\texttt{cc}}_{i,j} \mathbf{f}_{un, i} + \mathbf{b}_{i,j}^{\texttt{cc}}) \\
     \left(\sum_{n'}\text{ReLU}(\mathbf{W}^{\texttt{ca}}_{i,j} \mathbf{f}_{un', i} + \mathbf{b}_{i,j}^{\texttt{ca}})\right) \big/ N\\
     \left(\sum_{u'\neq u}\text{ReLU}(\mathbf{W}^{\texttt{oc}}_{i,j} \mathbf{f}_{u'n, i} + \mathbf{b}_{i,j}^{\texttt{oc}})\right) \big/ (U-1)\\
     \left(\sum_{u'\neq u}\sum_{n'}\text{ReLU}(\mathbf{W}^{\texttt{oa}}_{i,j} \mathbf{f}_{u'n', i} + \mathbf{b}_{i,j}^{\texttt{oa}})\right) \big/ (N(U-1)) 
\end{pmatrix},
\label{eq:expansion_layer_processing}
\end{multline}
where $j$ is the index of the \gls{ris} element in $\Lambda_i$.
Readers are referred to \cite{peng2025risnet} for more details.

\subsection{Joint Optimization of Precoding and RIS Configuration}

Problem~\eqref{eq:problem_sdma} is a joint problem of \gls{bs} precoding and \gls{ris} configuration.
While \gls{ris} configuration is a new problem and we propose to solve it with \gls{ml},
\gls{bs} precoding has been thoroughly studied in the literature.
Therefore,
we combine the analytical high-performance \gls{wmmse} precoder~\cite{shi2011iteratively}
and the \gls{ml} enabled \gls{ris} configuration as a hybrid solution.
In each iteration,
the channels are computed with fixed RISnet parameters,
based on which the precoding matrix $\mathbf{V}$ is computed.
A gradient ascent step is then performed to update the RISnet parameters
with fixed precoding matrices.
In this \gls{ao} approach,
we significantly reduce the learning difficulty
and guarantee the precoding performance.
\section{Training and Testing Results}
\label{sec:results}

\begin{figure}
    \centering
    \resizebox{.5\linewidth}{!}{		\begin{tikzpicture}[scale=1]
            \tikzstyle{base}=[isosceles triangle, draw, rotate=90, fill=gray!60, minimum size =0.12cm]
   
			\foreach \i in {1, 1.8, 3.4, 4.2, 5.0, 5.8}
			\foreach \j in {2, 4}
			{
				\fill[gray!30!white] (\i, \j) rectangle (\i+0.6, \j+1);
			}
   
			\foreach \i in {1.4, 3.4}
			\foreach \j in {-0.4, 0.8}
			{
				\fill[gray!30!white] (\i, \j) rectangle (\i+1, \j+1);
			}
			\node[align=center, rectangle,draw, minimum height=.65cm, text width=1cm] (ue) at (4.7,3.5) {Users};
			\draw (1,3.9) -- (2.5,3.9);
			\draw (1,3.1) -- (2.5,3.1);
			\draw (3.3,3.9) -- (6.4,3.9);
			\draw (3.3,3.1) -- (6.4,3.1);
			
			\draw (2.5,-0.4) -- (2.5,3.1);
			\draw (3.3,-0.4) -- (3.3,3.1);
			\draw (2.5,3.9) -- (2.5,5);
			\draw (3.3,3.9) -- (3.3,5);
			
            \node[base] (BS) at (2.45,0.1){};
			\node[below of=BS, yshift=.2cm] (bs) {BS};
            \draw[decoration=expanding waves,decorate] (BS) -- (3.2,1.9, 1.2);
			\node[rectangle, fill=white, draw, scale=.8] (ris) at (3.6,4.1)
			{  \hspace{.0cm}RIS\hspace{.0cm} };
	\end{tikzpicture}}
    \caption{The considered scenario: an intersection in an urban environment.}
    \label{fig:scenario}
\end{figure}
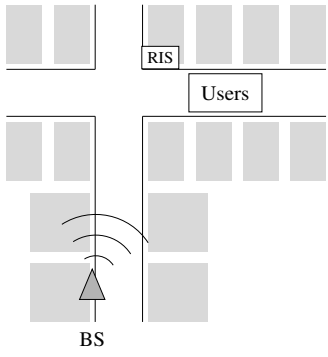

The open-source DeepMIMO dataset~\cite{Alkhateeb2019} is applied to generate the channel data.
The scenario is shown as \autoref{fig:scenario}.
User positions are randomly generated within the area shown in the figure.
Channel models are calculated according to the user positions.
The \gls{ris} has 1296 elements.
In total,
10240 samples are generated for training
and 1024 samples are generated for testing.
Samples in training and testing sets
are independent and identically distributed.

\begin{figure}[htbp]
    \centering
    \input{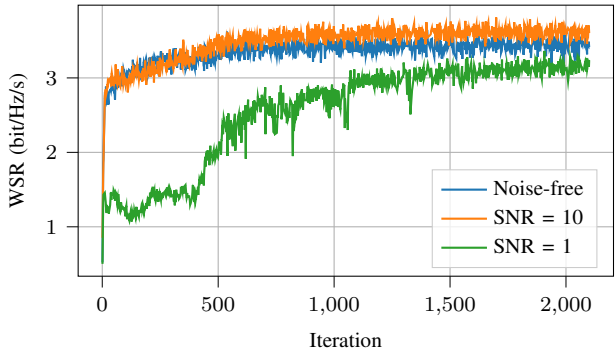}
    \caption{Realized WSR in training with different SNRs.
    As expected, lower SNR leads to worse WSR.
    }
    \label{fig:training}
\end{figure}

As explained in \autoref{sec:ml},
observation $\mathbf{O}$ is noisy due to thermal noise.
The \gls{snr} can be adapted via transmit power and pilot length.
In the following,
we train the RISnet with \gls{snr} of $\infty$ (i.e., no thermal noise), 10 and 1.
The training result is shown in \autoref{fig:training}.
It can be observed that the training significantly improves the \gls{wsr}.
While the \gls{snr} of 10 realizes a \gls{wsr} similar to the result without noise,
the \gls{snr} of 1 undergoes a more difficult training process.
This result invokes trade-off between resource allocation for channel estimation and
communication performance.
In addition to performance,
the trained RISnet requires 70~milliseconds to configure the \gls{ris},
whereas the \gls{bcd} algorithm requires 253~seconds
and the \gls{drl} model requires 0.23~second,
confirming the high efficiency of the proposed method.

\begin{figure}[htbp]
    \centering
    \input{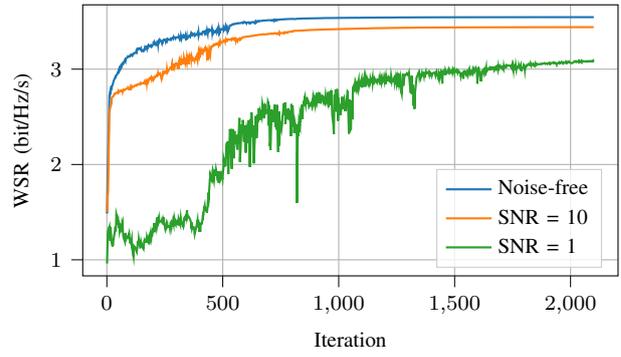}
    \caption{Realized WSR in testing with different SNRs.
    As expected, lower SNR leads to worse WSR.
    }
    \label{fig:testing}
\end{figure}

\autoref{fig:testing} shows the testing results with different \glspl{snr}.
We can observe a similar result as the training result.
Unlike training,
the \gls{wsr} without noise is higher than the \gls{wsr} with an \gls{snr} of 10,
suggesting a slight overfitting of the latter model.

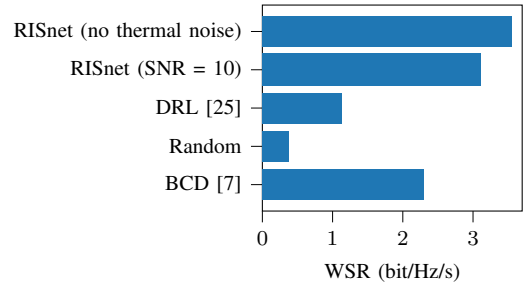
\begin{figure}
    \centering
\begin{tikzpicture}
\tikzstyle{every node}=[font=\footnotesize]

\definecolor{darkgray176}{RGB}{176,176,176}
\definecolor{steelblue31119180}{RGB}{31,119,180}
\tikzstyle{every node}=[font=\footnotesize]

\begin{axis}[
height=.5\linewidth,
tick align=outside,
tick pos=left,
x grid style={darkgray176},
xlabel={WSR (bit/Hz/s)},
xmin=0, xmax=3.7,
xtick style={color=black},
y grid style={darkgray176},
ymin=-0.69, ymax=4.69,
ytick style={color=black},
ytick={0,1,2,3,4,5, 6, 7},
yticklabels={
  BCD~\cite{guo2020weighted},
  Random,
  DRL~\cite{huang2020reconfigurable},
  {RISnet (SNR = 10)},
  {RISnet (no thermal noise)},
}
]
\draw[draw=none,fill=steelblue31119180] (axis cs:0,-0.4) rectangle (axis cs:2.30,0.4);  
\draw[draw=none,fill=steelblue31119180] (axis cs:0,0.6) rectangle (axis cs:0.382,1.4);  
\draw[draw=none,fill=steelblue31119180] (axis cs:0,1.6) rectangle (axis cs:1.13,2.4);  
\draw[draw=none,fill=steelblue31119180] (axis cs:0,2.6) rectangle (axis cs:3.11,3.44);  
\draw[draw=none,fill=steelblue31119180] (axis cs:0,3.6) rectangle (axis cs:3.55,4.4);  
\end{axis}

\end{tikzpicture}
    \caption{Performance comparison between proposed algorithm and baselines.}
    \label{fig:comparison}
\end{figure}

\autoref{fig:comparison} shows the comparison between proposed algorithm and baselines,
where a noise-free \gls{csi} is assumed for a fair comparison
because the baseline algorithms have not considered thermal noise.
It can be observed that the proposed algorithm realizes higher \glspl{wsr}
and it does not require explicit channel estimation for \gls{ris} configuration as baselines \gls{bcd}~\cite{guo2020weighted} and \gls{drl}~\cite{huang2020reconfigurable}.

\section{Conclusion}
\label{sec:conc}

The \gls{ris} is a promosing solution for the next-generation mobile radio systems.
However, it faces the key challenges of scalability and channel estimation.
In this work,
we have proposed a novel \gls{ml}-enabled \gls{ris} configuration to address both issues.
In particular,
we proposed an implicit channel estimation scheme to drop the unrealistic assumption of known \gls{csi}.
Moreover,
a dedicated \gls{nn} architecture RISnet was proposed,
which is able to configure a large \gls{ris} with more than 1000 elements.
Simulation results showed that there exists a trade-off between resource allocation for implicit channel estimation
and communication performance.
Furthermore,
the proposed algorithm outperformed baseline algorithms significantly.

Code and data of this work are available under \url{https://github.com/bilepeng/risnet_implicit_channel_estimation}.

\printbibliography

\end{document}